\documentclass[fleqn,10pt]{wlscirep}
\usepackage[utf8]{inputenc}
\usepackage[T1]{fontenc}
\title{Predicting System Dynamics of Universal Growth Patterns in Complex Systems}

\author[1,*]{Leila Hedayatifar}
\author[1,+]{Alfredo J. Morales}
\author[1]{Dominic E. Saadi}
\author[1]{Rachel A. Rigg}
\author[1]{Olha Buchel}
\author[1]{Amir Akhavan}
\author[2]{Egemen Sert}
\author[1]{Aabir Abubaker Kar}
\author[3]{Mehrzad Sasanpour}
\author[1,4]{Irving R. Epstein}
\author[1,*]{Yaneer Bar-Yam}
\affil[1]{New England Complex Systems Institute, 125 Mount Auburn St., Box 380762, Cambridge, MA 02138, USA}
\affil[2]{Department of Computer Engineering, Middle East Technical University, Ankara, Türkiye}
\affil[3]{Department of Physics, Sharif University of Technology, Tehran, Iran}
\affil[4]{Department of Chemistry, Brandeis University, Waltham, MA 02453, USA}

\affil[*]{leila@necsi.edu, yaneer@necsi.edu}

\affil[+]{Deceased}


\begin{abstract}
Predicting dynamic behaviors is one of the goals of science in general as well as essential to many specific applications of human knowledge to real world systems. Here we introduce an analytic approach using the sigmoid growth curve to model the dynamics of individual entities within complex systems. Despite the challenges posed by nonlinearity and unpredictability in system behaviors, we demonstrate the applicability of the sigmoid curve to capture the acceleration and deceleration of growth, predicting an entity's ultimate state well in advance of reaching it. We show that our analysis can be applied to diverse systems where entities exhibit nonlinear growth using case studies of (1) customer purchasing and (2) U.S. legislation adoption. This showcases the ability to forecast months to years ahead of time, providing valuable insights for business leaders and policymakers. Moreover, our characterization of individual component dynamics offers a framework to reveal the aggregate behavior of the entire system. We introduce a classification of entities based upon similar lifepaths. This study contributes to the understanding of complex system behaviors, offering a practical tool for prediction and system behavior insight that can inform strategic decision making in multiple domains.
\end{abstract}

\begin{document}

\flushbottom
\maketitle

\thispagestyle{empty}
\section*{Introduction}

Nonlinearity often appears in systems of interacting components as they increase in size and complexity \cite{schoukens2019nonlinear, herrera2020tractable}. This phenomenon is observed in various real world systems, including social \cite{hedayatifar2019us}, economic \cite{sert2020freight}, and biological \cite{ozccelik2023overcoming,cohen2022complex} systems, where intricate nonlinear interactions introduce a high degree of complexity \cite{may1976simple}. In systems comprised of a multitude of interacting elements, the behavior often becomes irregular and challenging to predict. For instance, in economic markets, individual stock values may exhibit complex fluctuating patterns of behavior, and the aggregate of these patterns can give rise to collective booms and busts. These patterns, observed in a multitude of nonlinear systems across contexts or scales, challenge our ability to develop an understanding of system behavior, but also hold the potential to lead to profound insights \cite{jahanshahi2021development}. Such insights are crucial because recognizing and understanding patterns across systems can lead to enhanced predictability of system behavior, more effective system design, and the development of improved control strategies \cite{wabersich2021predictive}.
 
The emergence of dynamic complexity can be attributed to interactions among a system's constituent elements and/or to the nonlinear behaviors exhibited by these elements\cite{liang2021measuring, bar2019dynamics}. Such complexities typically give rise to nonlinear collective behaviors that defy direct proportionality to the individual behaviors of the entities, thus imparting unique characteristics to the system \cite{bak2021stewardship, higgins2002nonlinear, liu2016control}. In these systems, the conventional characterization of input-output relationships does not hold \cite{chia1988geometrically}. The central challenge for system description lies in identifying universal properties so that it is not necessary to develop distinct characterizations for each individual element or system \cite{tang2020introduction, barabasi1999emergence, may1976simple}. While universality may arise  at the aggregate level due to the effects of interactions among elements, it can also manifest at the individual level when the behaviors of individual entities can be succinctly described by a parameterized mathematical function \cite{strogatz2018nonlinear}. 

Many investigations into nonlinear systems have focused on the premise that individual entities of the system adhere to specific patterns \cite{hwa2004universal}. When there is noise in the dynamics of individual entities, predicting their behaviors over the short term may become challenging. However, their longer term dynamics may be still be well characterized by statistical dynamic functions \cite{feigenbaum1989universal, zheng2022data}, especially when the aggregate behavior of the system does not exhibit abrupt shifts \cite{feigenbaum1989universal}. Studying the dynamics of all entities of a system can unveil emergent patterns that might not be evident when examining individual entities in isolation. These patterns can assume various forms, from predictability to chaos, simplicity to complexity, reflecting the inherent nature of such systems \cite{vulpiani2009chaos}. 

In this study, we have developed an analytic  approach aimed at investigating dynamical systems, focusing on systems where the individual entities exhibit specific nonlinear growth behaviors over time. In these systems, an entity's activity grows and then diminishes, exhibiting phases of acceleration and deceleration, akin to a bell-shaped curve. The initiation of activity sets in motion a self-activating pattern that increases activity over time. However, due to various internal or external constraints, an inhibitory pattern eventually arises, slowing down the activity, eventually leading the entity to end its activity. The activating and inhibiting processes give rise to unique behaviors for each entity that include stochastic variation. To analyze and understand such behaviors, we employ a sigmoid function, which we demonstrate to be effective as a universal description of the aggregate over time of growth and decline. This nonlinear function \cite{Richards1959, carrillo2002new} captures phenomena that commence gradually, undergo acceleration, and eventually reach saturation, creating an "S"-shaped pattern. Sigmoid functions have previously been utilized to model diverse phenomena, ranging from the growth of biological populations \cite{Thieme2003} and the spread of infectious diseases \cite{Hsieh2006} to the adoption of new consumer innovations \cite{sultan1990}. Leveraging real world historical data, a sigmoid analysis allows us to predict both the time when an entity is likely to saturate (ending activity) and the total amount of its activity. The parameterized trajectories for each entity can be visualized, enabling us to gain insights into its lifepath, providing both quantitative prediction for the near future and characterization of the long term behavior. 

We utilize two distinct datasets. The first is a database of industrial customer orders. While customer purchasing behavior is inherently complex, as each customer seeks to satisfy individual needs \cite{Pennacchioli2014}, we show that the trajectory of order frequencies follows a universal behavior consistent with sigmoid functions. Customers place initial orders and progressively increase their frequency until they gradually reduce their orders and eventually stop. This behavior may reflect changes in market conditions, competition, time and/or resource constraints that limit customer purchasing and give rise to the observed nonlinear behavior. The second dataset pertains to legislative process, including introduction of new bills by parliaments, congresses, or other similar institutions. We use texts of proposed laws to extract named entities about acts, environmental agencies, associations, agencies, boards, various codes delineating minimum requirements (e.g., building codes or energy conservation codes), and numerous other named entities. The dynamics of the usage of names of such entities in introduced bills reveal a nonlinear behavior that can similarly be characterized by sigmoid curves. We show that awareness of the dynamics of customer orders and term usage in legislation offers valuable predictive tools for these contexts.

\section*{Results and Discussion}

A sigmoidal curve is defined by three key parameters: inflection point, slope, and amplitude. Here, we show that the evolution of the fitted parameters provides a clear understanding of the current state of activity and can make reliable predictions about future behavior. The results for the primary dataset are presented in this section, while those for the second dataset are available in the supplementary materials. In Figure \ref{fig1}A, we show a sigmoid curve fit to the cumulative time series of a particular customer who initiated their orders in 2006 and ceased ordering in 2008. The customer transitioned from an accelerating phase to a decelerating one early in 2007. The model suggests that the customer is unlikely to place any new orders after 2008. Using the reduced $\chi^2$ statistic (Fig. \ref{fig1}B), we find the sigmoidal fit significantly outperforms a linear fit, as seen in smaller $\chi^2$ values. In Fig. \ref{fig1}C, the sigmoid fit and its parameters are shown for consecutive years for another representative customer. We see that as a customer continues to grow with new orders over the years, the parameters of its sigmoid fit change. In the following, we show how by tracking the sigmoid parameters we can analyze their behavior and predict future trends.

\begin{figure*}[ht!]
\centering
\includegraphics[width=0.9\textwidth]{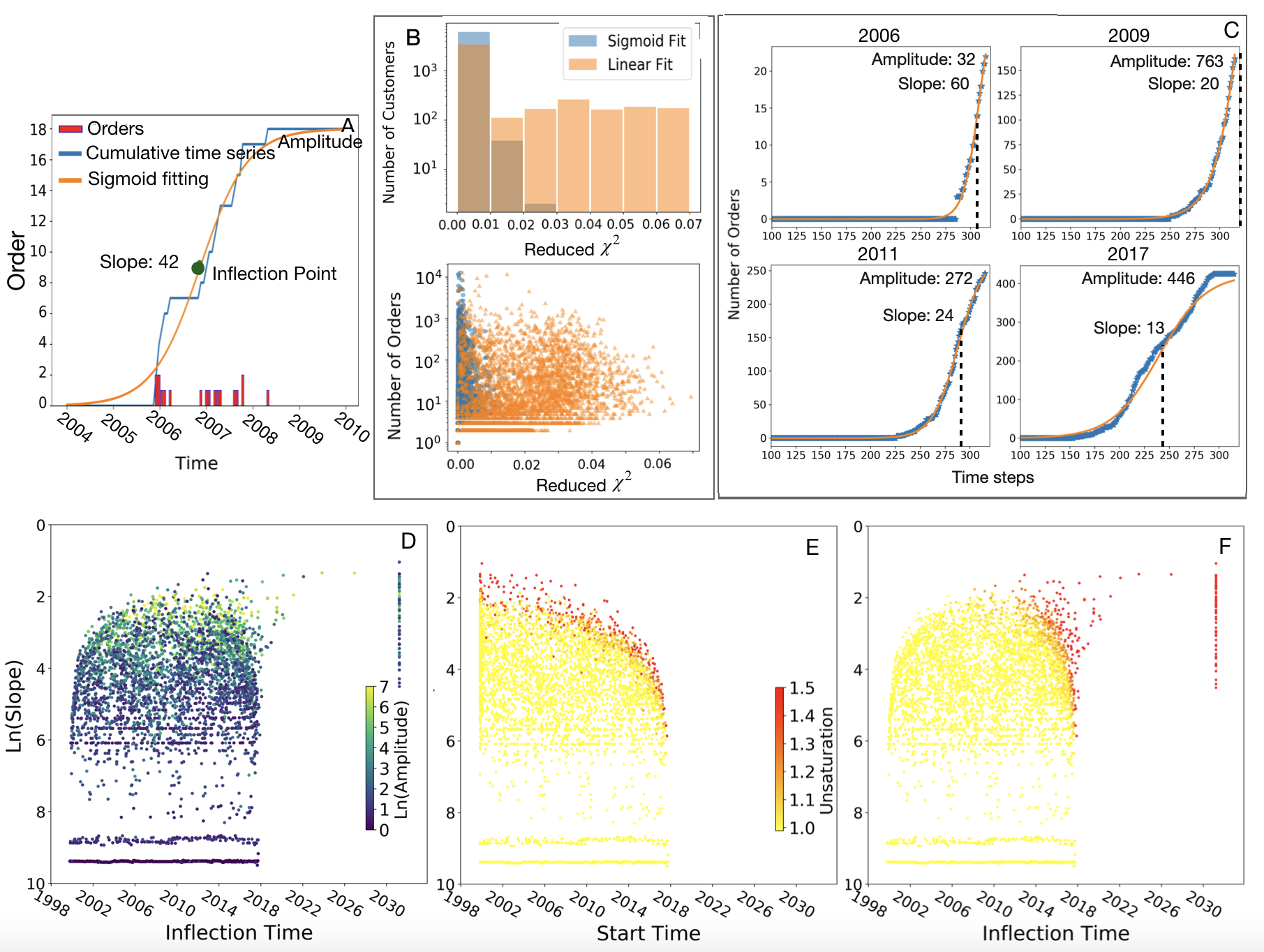}
\caption{\label{fig1}Sigmoid curve fitting and parameter spaces for customer ordering behavior. (A) Customer time series and fitted sigmoid curve. Red bars represent the number of orders over time for a customer; the blue line is the cumulative time series; and the orange line is the sigmoid curve fitted to the cumulative time series. Sigmoid curve parameters include inflection time (green dot, center), slope of the curve at the inflection time, and amplitude or total orders. (B) Comparison of the linear and sigmoid fits to the customers using the reduced $\chi^2$ test. (C) Ordering behavior of a representative customer in different years and the fitted sigmoid curves and their parameters. Parameter spaces for all customers are shown for (D) slope, inflection time, and amplitude, (E) slope, start time, and saturation status, and (F) slope, inflection time, and saturation. The third parameter is shown in color (scale in inset). For the customers that are early in their growth, the inflection time and amplitude are not reliable, so we assign them an arbitrary inflection time of 2030 with their last total orders as amplitude.}
\end{figure*}

To visualize properties of the entire population, we constructed parameter spaces, where dots represent the coordinates of each entity for three parameter values, with color serving as the third dimension. These parameter spaces reveal collective system properties that are not visible when examining individual entity behaviors. In addition to the three primary parameters---inflection time, slope, and amplitude---two other pertinent parameters are the start time of activity and the saturation value, defined as the ratio of the fitted amplitude $A$ to the final number of cumulative orders. In the parameter space defined by inflection time, slope, and amplitude (Fig. \ref{fig1} D), customers who place orders over an extended duration tend to exhibit a lower slope and are situated in the upper area of this parameter space. Repeat customers, marked by green-yellow dots, occupy this region as they consistently place orders over an extended period. Additionally, a few smaller customers in the purple category also appear in this segment when their orders span a considerable time frame. Conversely, customers who make a single order are characterized by the steepest slopes and are represented as purple dots positioned at the bottom of the graph. In cases where customers are at the nascent stage of their growth and exhibit a steep rise in slope, the inflection time becomes an unreliable metric, and is given an arbitrary value of 2030. In our model, a saturation status approaching 1.0 signifies the point at which customers cease placing orders, while values exceeding 1.0 indicate a likelihood of future orders. In the parameter space defined by start time, slope, and saturation, we observe that customers with lower slopes at any given start time are categorized as unsaturated (represented by red dots), indicating their potential for continued growth (see Fig. \ref{fig1} E). Furthermore, inflection times for unsaturated customers are either in the recent past, meaning the rate of ordering has begun to slow down, or in the future, indicating customers with the most growth potential (Fig. \ref{fig1} F).

\begin{figure*}
\centering
\includegraphics[width=0.95\textwidth]{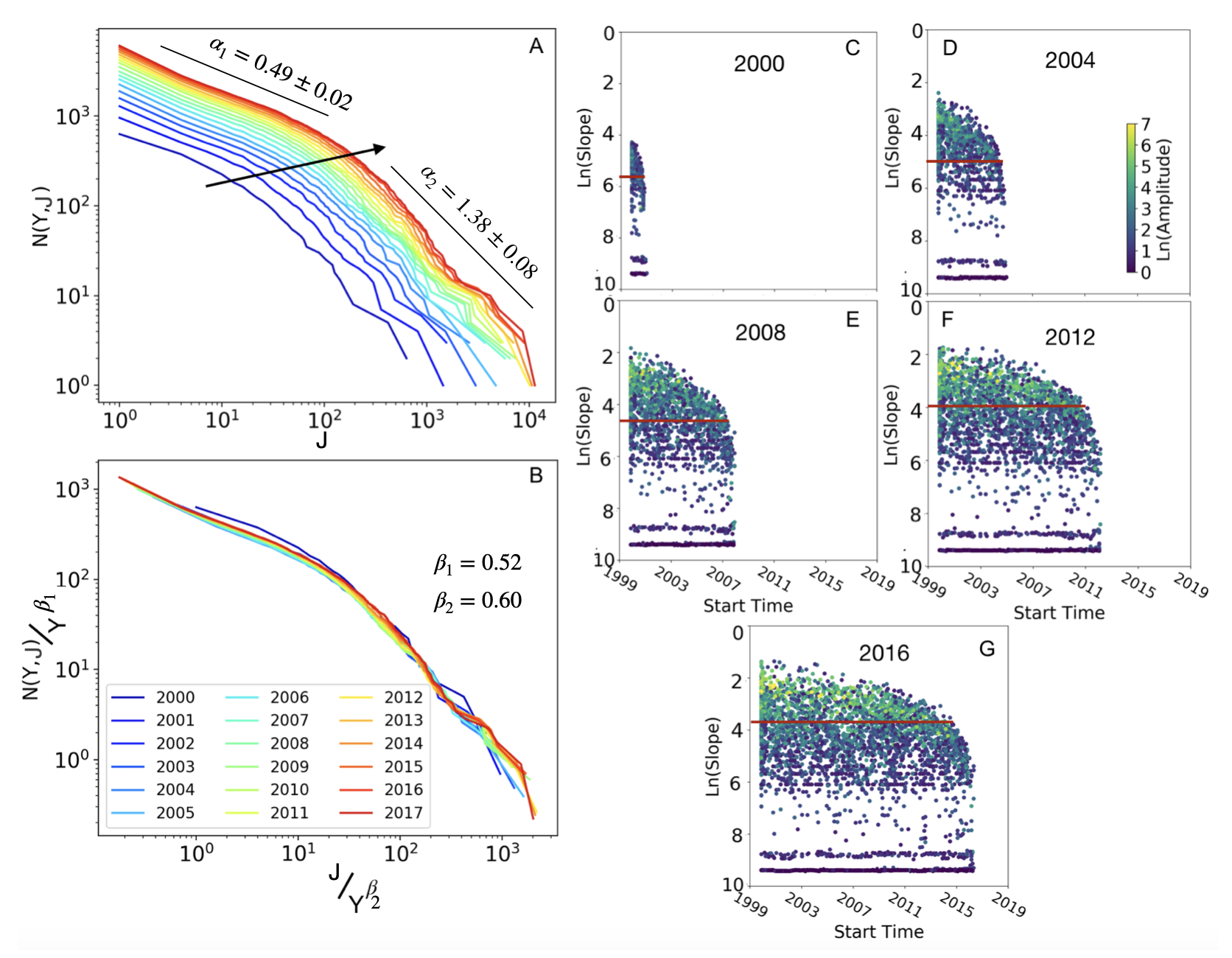}
\caption{\label{fig2}Universality in customer ordering behavior illustrated by the progression of parameter distributions over time. (A) Number of customers $N(Y,J)$ that place more than $J$ orders over different periods from $1999$ to year $Y$ (colors, legend in (B)). The black arrow represents the change in the breakpoints over the years. (B) Rescaling of the $J$-axis and $N$-axis by $Y^{\beta_1}$ and $Y^{\beta_2}$, respectively. Parameter spaces for customer slope, start time, and amplitude are shown for the periods from 1999 to the years (C) 2000, (D) 2004, (E) 2008, (F) 2012, and (G) 2016, with amplitude in colors (key, upper right). The red line on each graph represents the slope at the breakpoint in aggregate distributions from (A).} 
\end{figure*}

Shifts and fluctuations in the distribution of the target variable pose challenges for predicting the collective behavior of a system. In contrast, when the entities within a system exhibit consistent behavior, it enhances our ability to make more reliable predictions, both at the collective level and for individual behaviors \cite{bar2019dynamics, turner1957collective, goldstone2009collective}. In this regard, power law distributions hold great significance, because they unveil a fundamental regularity in system properties at various scales. In a system with a power law distribution, transitions between phenomena at various scales remain consistent regardless of the specific scale under consideration. This self-similarity forms the foundation of power law relationships \cite{jafari2007long, gallos2007review}. Analyzing the distribution of entities based on their key parameters over various time durations provides a valuable approach for assessing the consistency in system dynamics. By plotting these distributions and observing their trends over time, fitting a power law can serve as an effective method to describe the system's enduring patterns and stability. 

In our customer dataset, we charted the distributions of customers based on their order patterns over varying timeframes, spanning from 1999 through the subsequent years. (Fig. \ref{fig2} A). The distributions count the number of customers $N(J)$ that place at least $J$ orders. The distributions follow a power-law $N(J)\sim J^{-\alpha}$. They include two regimes, with different $\alpha$ exponents. The regimes and the breakpoint between them shift toward larger numbers of customers $N$ and orders $J$ over time, as shown by the black arrow. In the first regime, $\alpha_1 = 0.49\pm 0.02$, and in the second regime, $\alpha_2 = 1.38\pm 0.08$, both of which are consistent for all years. The difference in the exponent of the two regimes reveals that the proportion of larger customers decreases more slowly in the first regime than in the second regime. If $Y$ represents the number of years we aggregate the data, i.e., $Y=1,2,3,...,18$, then by scaling the axes by a specific power of years, respectively, all the curves collapse into a single curve described by $N(Y,J)/Y^{\beta_1} \sim f(J/Y^{\beta_2})$, with $\beta_1=0.52$ and $\beta_2=0.60$ (Fig. \ref{fig2}B). The scaling function $f(u)$ demonstrates that the customer distribution is universal and can be generalized to future years. 

The rationale behind the growth in these distributions and the identification of breakpoints becomes apparent when we observe the evolution of the parameter space encompassing start time, slope, and amplitude across the years. (Fig. \ref{fig2} C-G). Over time, as the number of customers and their lifespans grow, we observe an increase in the quantity of points within the parameter space, which shift towards smaller slopes and larger amplitudes. To identify the average slope of customers at the breakpoints (red lines in Fig. \ref{fig2} C-G), we leverage the joint distribution of slopes and amplitudes. From figure 1E, one can find that active customers (red dots) are in the upper portion of the parameter space across all years. Slopes with values less than that at the breakpoint (i.e., points above the red line in Fig. \ref{fig2} C-G) encompass customers who have been active across all years. Conversely, larger slopes reveal a gradual decline in customer numbers in recent years, primarily because the most recent customers have not had sufficient time to accumulate numerous orders. This disparity in the exponents of the power law behavior underscores the distinction between the two regimes. As the red line progressively shifts upwards over time, it captures an increasing number of customers with higher order counts situated beneath it. This signifies a burgeoning in the first regime of the distributions, reflecting the expansion of overall business activity.

At each time $t$ (year), we calculated the probability of having a customer that starts ordering at time $t_0$ and reaches ln(slope) $m'$ with ln(amplitude) $A'$, $p(t_0,m',A')$. See Supplemental Materials for details and Figures. The dependence of our equation on $t$ is hidden in the start time, $t_0$, parameter. As the slope and amplitude parameters are not independent, we define the probability as,
\begin{equation}
    p(t_0,m',A') =p(A'|m',t_0)p(m'|t_0)p(t_0).
    \label{factorization}
\end{equation}
We found that customer start times, $t_0$, are uniformly distributed, i.e., 
\begin{equation}
    p(t_0) =\begin{cases}
      \frac{1}{t_{0b}-t_{0a}} &  t_{0a}\le t_0\le t_{0b} \\
      0 & otherwise
    \end{cases}
    \label{start-time}
\end{equation}
where $t_{0a} = 0.32$ and $t_{0b} = 1$ represent the first and last months in scaled cumulative time series at $t=2017$. The value of $t_{0b}$ is always equal to 1, and $t_{0a}$ depends on the time $t$ at which we do the analysis. While the probability of having large slopes (small customers) is independent of their start time, we find that for smaller slopes it is a function of $t_0$, $m'_a = 0.5+e^{1.5t_0^{4}}$ (see Figure S4A). Thus 
\begin{equation}
    p(m'|t_0) =\begin{cases}
      \frac{0.57}{6.5-m'_{a}}  &  m'_{a}<m'<6.5 \\
\frac{0.03}{9.1-8.9} & 8.9<m'<9.1\\
      \frac{0.40}{9.4-9.3} & 9.3<m'<9.4.
    \end{cases}
    \label{slope-start-time}
\end{equation}
We did not attempt to fit the data in the ranges of ln(slope), $6.5 < m' < 8.9$ and $9.1 < m' < 9.3$, where there were too few points to permit a reliable assessment of the distribution.
We observe that the ln(amplitude) $A'$ displays a normal distribution as a function of  ln(slope) $m'$, as
\begin{equation}
    p(A'|m',t_0) = \begin{cases}
      \frac{1}{\sqrt{2\pi} \sigma_1}e^{-\frac{(A'-\mu_1)^2}{2\sigma_1^2}} &  m'_{a}<m'\le 6.5 \\
\frac{1}{\sqrt{2\pi} \sigma_2}e^{-\frac{(A'-\mu_2)^2}{2\sigma_2^2}} & 6.5<m'\le 9.4
    \end{cases}
    \label{amplitudeslopeeq}
\end{equation}
where the parameters for the first regime are given by $\mu_1=a_0m'^2+a_1m'+a_2$ and $\sigma_1=b_0m'+b_1$, with $a_0=0.13$, $a_1=-2.12$, $a_2=9.86$, $b_0=-0.20$, and $b_1=1.93$, and the parameters for the second regime are $\mu_2=1.0$ and $\sigma_2=\ln(1.5)$. Note that the dependence of this equation on $t_0$ is hidden in the value of $m'_a$.

\begin{figure*}
\centering
\includegraphics[width=0.9\textwidth]{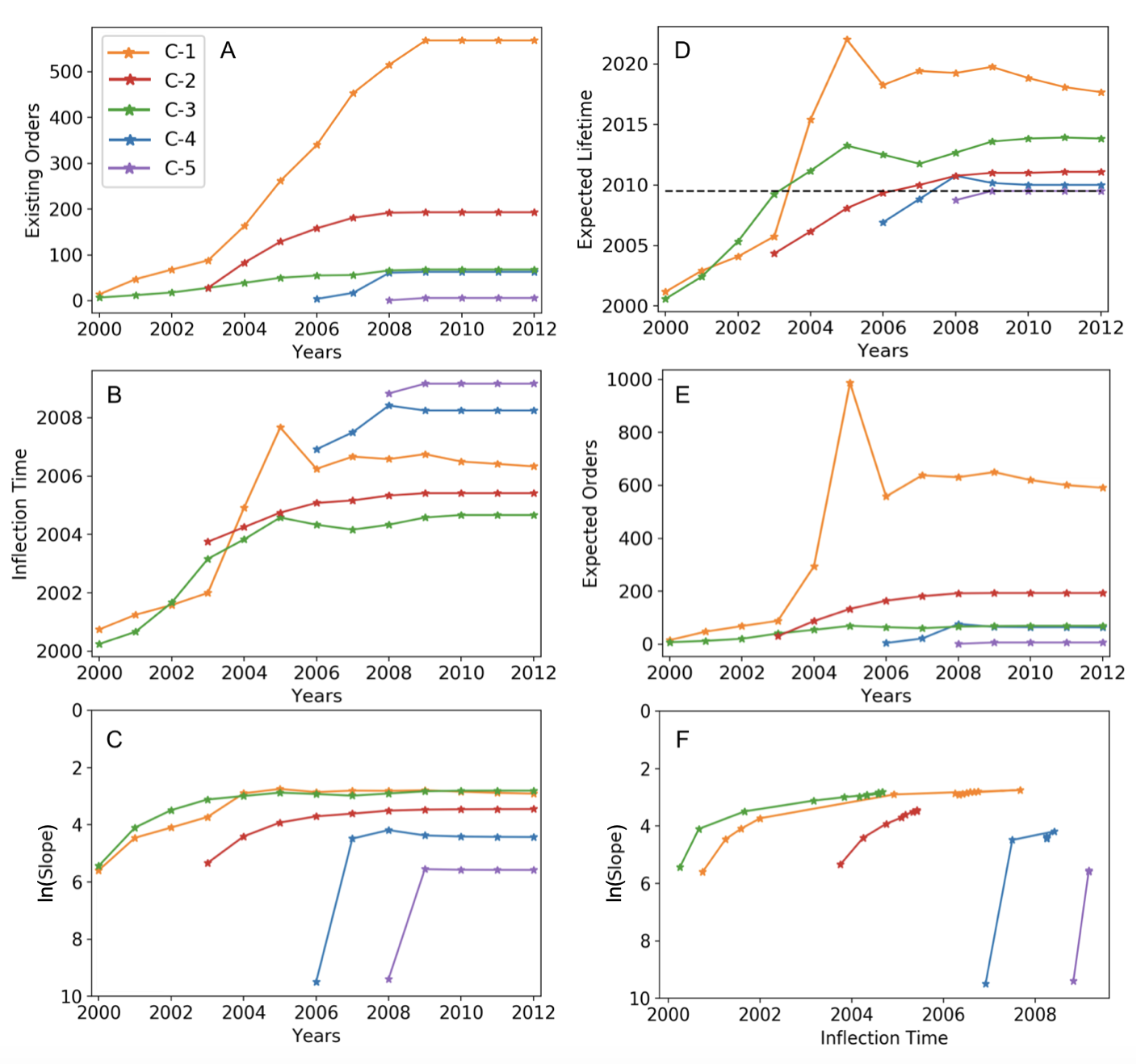}
\caption{\label{fig3}Lifepaths of five representative customers. Each customer (legend, inset in (A)) started in a particular year and left the system in 2009. (A) Cumulative orders per year. Sigmoid parameters are calculated each year from the monthly time series of orders, including (B) inflection time, (C) ln(slope), (D) expected customer leaving time, and (E) expected total orders. (F) ln(slope) versus inflection time; closeness of points indicates a customer has already passed their final inflection time.}
\end{figure*}

The dynamics of the sigmoid curve parameters enable prediction of future system dynamics. We can assess the predictive accuracy of the sigmoid fit by selecting already saturated entities and examining their gradual changes. Furthermore, we can analyze how successfully the sigmoid parameters predict their behavior. 

To illustrate visually the activity dynamics of entities, we chose five representative customers that entered the system at different times and left the system in 2009, and we tracked the evolution of their sigmoid parameters from prior years to their "final" values in 2009 and beyond (Fig. \ref{fig3}). The five selected customers include large customers who placed many orders each year (C-1 and C-2), medium-sized customers who placed a few orders each year (C-3 and C-4), and a small customer who placed a few orders in one year (C-5) (Fig \ref{fig3} A). We examined the evolution of inflection time (Fig. \ref{fig3} A) and slope (Fig. \ref{fig3} B). Because each customer left the system in 2009, the sigmoid parameters in 2009 and afterwards represent the aggregate behavior of the customer. Until the final value of the inflection time is reached, the predicted inflection time moves forward to later times and the slope tends to smaller slopes.  After passing the final inflection time, the sigmoid function predicts the same parameters from then on, i.e, inflection time and slope do not change after that year. Therefore, the stabilization of parameters is a sign that entity activity has passed the final inflection time. We also observe a stabilization in the evolution of expected customer leaving time (Fig. \ref{fig3} D) and expected number of orders (Fig. \ref{fig3} E). However, for some customers (see C-1), the sigmoid fit shows a peak in inflection time (Fig. \ref{fig3} B), expected leaving time (Fig. \ref{fig3} D), and expected total orders (Fig. \ref{fig3} E) one year before the final inflection time. This occurs for customers that are growing rapidly. Furthermore, we can track the evolution of each customer as a trajectory through the parameter space of inflection time and slope (Fig. \ref{fig3} F). In these trajectories, large distances between consecutive points indicate a customer is still in the first regime of the sigmoid curve (before the final inflection time), while closeness of the points reveals the entity has passed the inflection time and will leave the system in the near future. If the ordering behavior of an entity changes for a period of time or the entity activity ceases temporarily, the trajectory moves backwards and forwards, providing a signal that the entity behavior may warrant further attention.

\begin{figure*}[t!]
\centering
\includegraphics[width=1\textwidth]{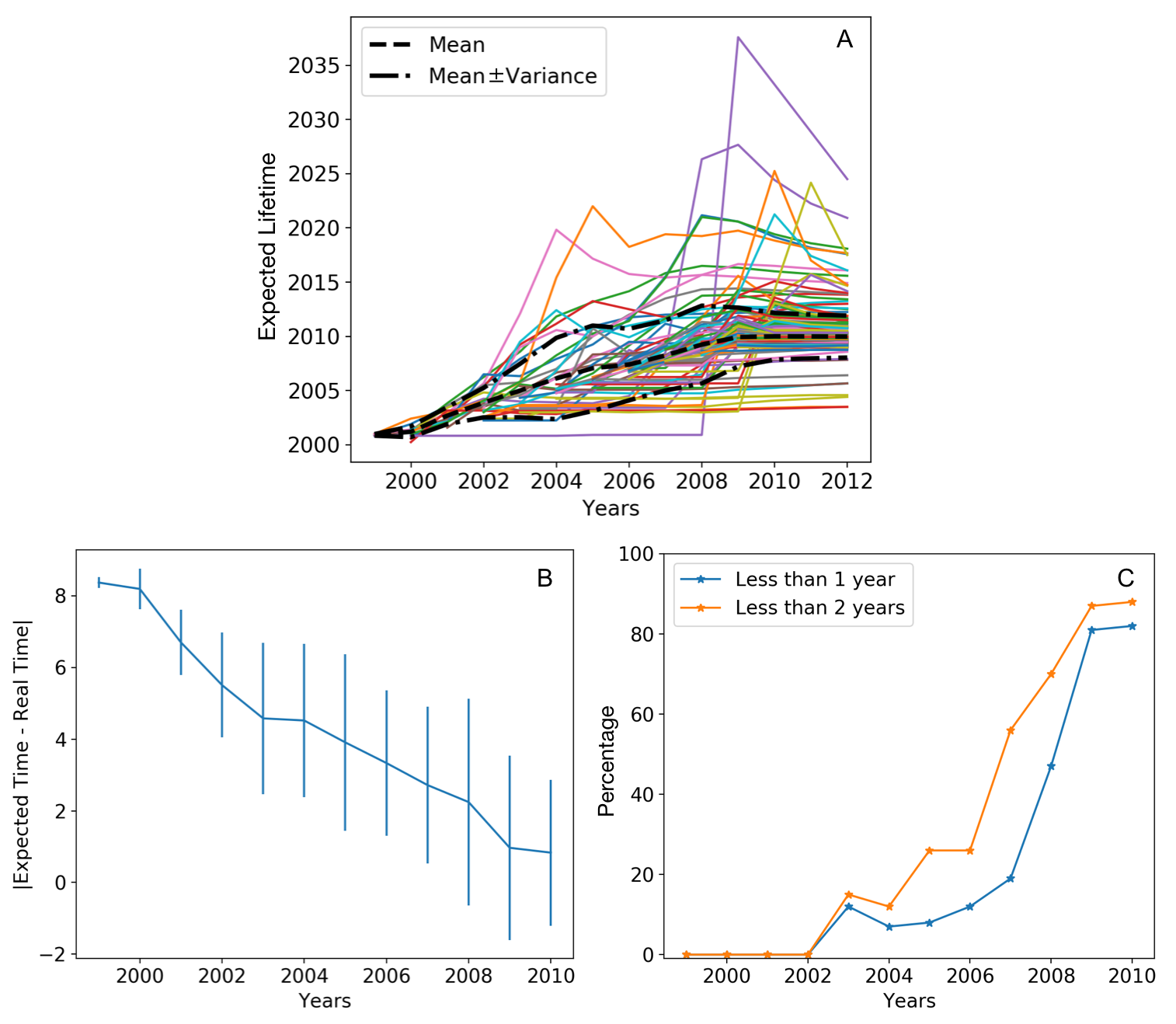}
\caption{\label{fig4}Validation of sigmoid fit predictions. Selected customers are those who entered the system in various years but all left in 2009. (A) Year a customer is expected to leave, according to the sigmoid fit, per year. Colored lines represent customers. The black dashed line represents the mean for all customers, and the black dash-dotted lines above and below represent the mean plus or minus the variance, respectively. Expected years should converge to 2009, the year all selected customers left. (B) Mean of expected time minus real time per year; error bars are variance. (C) Percentage of entities whose leaving time is predicted each year within one-year (blue line) or two-year accuracy (orange line).}
\end{figure*}

To validate predictive performance of the sigmoid fit, we picked a sample from the customers that had already left the manufacturing company. We selected the 364 customers who entered the system at any time but left in 2009. We examined the evolution of the expected leaving time of each customer calculated from the sigmoid curve fittings (Fig. \ref{fig4} A). Each colored line represents the expected leaving time of a single customer, and the black dashed line and dot-dashed lines show the mean and mean$\pm$variance, respectively. While variance is very small in the first few years, it increases to around 2-3 years in the years approaching 2009. Fig. \ref{fig4} B shows the mean of the difference between the expected time and real time in each year; error bars are the variance. In the first few years, when the history of orders is short, the expected leaving time is far earlier than 2009, and the mean difference is around 8 years. As we get closer, the mean leaving time increases towards 2009, and two years before 2009, the mean difference becomes less than a year. Our method is able to predict the leaving time for 56\%  of customers within a one-year difference in 2008 and for 82\% in 2009, and within a two-year difference for 58\% in 2007, 74\% in 2008 and 85\% in 2009 (Fig. \ref{fig4} C). Overall, we find the sigmoid fit predictions are accurate, with the exception of some fast-growing large entities, which the sigmoid fitting overevaluates and gives a chance to return, and long-lived small entities, which the sigmoid fitting underevaluates and assumes have already left.

\section*{Methods}
\subsection*{Data}

We conducted an analysis of entity behaviors in two distinct datasets. The first dataset, from a medium-sized U.S. manufacturing company, recorded orders from 6,065 customers during the time period spanning from November 1999 to September 2017. The number of orders per customer varies from 1 to over 10,000 orders over a period of 18 years. This dataset captures the company's customer interactions from the moment they initiated data recording, even though the company had already established a customer base prior to this period. Each customer is identified with a unique customer ID. The data include the year, month, day, minute and second of each order. Our analysis is not sensitive to smaller time scales so we aggregated orders by month. 

The second data set describes legislative developments related to per- and polyfluorinated substances (PFAS). Commencing from the 2016-2022 period, United States (US) federal and state legislatures have enacted numerous bills prohibiting the use of PFAS in a wide array of products. LegiScan, a legislative tracking service facilitating the monitoring and analysis of legislation at both the state and federal levels in the US, provides extensive data for legislative events. We collected all bills related to PFAS from LegiScan. We used queries such as PFAS OR PFOA OR polyfluor OR perfluor OR ethylene oxide OR microfiber OR microplastic OR halogenated flame retardant OR organohalogen. The results included 228 federal and state bills. We extracted the complete texts of the most recent versions of these bills. Employing natural language processing, we extracted the most significant named entities from the text, and high-frequency keywords frequently associated with PFAS terminology (e.g., product names, action verbs). In total, we identified 3100 relevant terms and their usage.

\subsection*{Sigmoid model and universal patterns}

To elucidate the behavior of each entity, $i$, we generated time series, $x_i(t)$, representing the monthly activity (purchasing volume, term presence) of the entity over time. Then, we generated the cumulative time series, $y_i(t)=\Sigma_{t_{0i}}^{t} x_i(t)$, where $t_{0i}$ represents the first activity for entity i.  While the specific interactive patterns of entities may differ based on particular conditions, a common underlying behavior emerges. Entities initiate their activities at specific times, gradually increase activity over subsequent periods, and then taper off until they ultimately cease activity. This nonlinear pattern---a slow start, acceleration, and deceleration to an end, forms an "S"-shaped curve, as described by Han (1995) \cite{Han:1995}. We fit this behavior to a sigmoid curve defined by three key non-negative parameters: the slope (m), the inflection point ($t_0$), and the amplitude (A) in the  equation: 
\begin{equation}
    y(t) = \frac{A}{1+e^{-m(t-t_0)}}
\end{equation}
The slope parameter corresponds to the gradient of the sigmoid at the inflection time, akin to  Bolton's concept of relationship depth in the marketing literature \cite{m1}. The inflection time parameter is the moment in time when the activity transitions from acceleration to deceleration. The amplitude parameter $A$ describes or anticipates the total activity by the entity over all time. 

The fitting of sigmoidal curves is sensitive to the length of the cumulative time series when the time series is relatively short. To ensure consistency in the fitting process for entities of varying lifespans, we standardized the length of all cumulative series. This standardization involved augmenting each time series with an array of zero values at times prior to the first recorded activity. We added 100 months of zero values at the beginning of each time series. We then rescaled the cumulative time series and $t$-axes to a range from 0 to 1, enabling comparisons between sigmoid curves and their respective parameters. Entities that initiated activity within the final two time points were excluded from the analysis due to insufficient data. Following these adjustments, we fit a sigmoid curve to the scaled cumulative curve. If the sigmoid curve exhibits an amplitude less than or equal to 1.0, it indicates saturation, suggesting that the entity has either ceased interactions within the observed time frame or will do so imminently. Conversely, an amplitude greater than 1.0 signifies an unsaturated curve, indicating that the entity is likely to undergo more activity in the future. After the fitting procedure, the results are rescaled back to their actual values.

Analyzing the evolution of parameters with a variable ending time of the analysis provides valuable insights into the patterns of behavior and their universality. The universality allows us to predict what will happen for the components and the system in the future. To examine the distribution of entities versus their activities, we employ a survival function. This function quantifies the number of entities with a cumulative activity of at least $J$, denoted as $N(J)$, for the range $J = {1, 2, 3, ..., J_{max}}$, and is defined as $N(J) = \sum_{i=J}^{J_{max}} n_i$, where $n_i$ is the number of entities with cumulative activity $i$. Our investigation characterizes the distribution of cumulative activity, referred to as entity size. The shape of this distribution reflects intrinsic properties of the system and may evolve over time, as indicated by previous studies \cite{Benguigui, Mitzenmacher2004}. In cases where the system encompasses a small number of entities with substantial activity levels and a multitude of small entities (i.e., with small levels of activity), the distribution exhibits $skewness$, or fat tailed properties. One extreme skewed distribution is a power law distribution \cite{Newman2005a, Bryson1974}. Power law distributions are described by the expression $N(J)\sim J^{-\alpha}$, where $\alpha$ serves as a measure of the degree of fat-tailed behavior. 

In conclusion, the sigmoid curve has been used to model the nonlinear growth of systems with a limited leaving time. Our analysis shows that a sigmoid curve can usefully model the dynamics and behavior of individual entities within these systems. Our model is applicable for any system in which entities initiate, accelerate (engaging phase), decelerate (disengaging phase) and finally cease their activity. Depending on the distance of the inflection time from the start time and the scale on which we look at the system, this model can predict the stop time of an entity far in advance. Studies of a customer system and the US legislative system, using the sigmoid model can predict the stop time of a customer or of a legal term multiple months, up to years in advance. This information can help analysts and policy makers distinguish the behavior of entities and develop strategies to enhance their policies. Furthermore, the parameter spaces generated from entity behavior provide a framework that describes properties of the aggregate behavior of the system.

\bibliography{apssample}

\section*{Acknowledgements}

We acknowledge the invaluable contributions of our coauthor, Alfredo Morales, who passed away during the preparation of this manuscript. His expertise and insights were critical to the development of this work and we dedicate this paper to his memory. The authors thank William Glenney for feedback and Matthew Hardcastle for proofreading an earlier version of the manuscript.

\section*{Author contributions statement}

All authors contributed to collecting data and analyzing the results. L.H. and R.A.R wrote the manuscript. L.H., Y.B., and I.R.E reviewed the manuscript. 

\end{document}


\maketitle
\renewcommand{\abstractname}{} 
\renewcommand{\abstract}{}

\section*{I. Customer distributions}
\begin{figure}[ht!]
\centering
\includegraphics[width=0.7\linewidth]{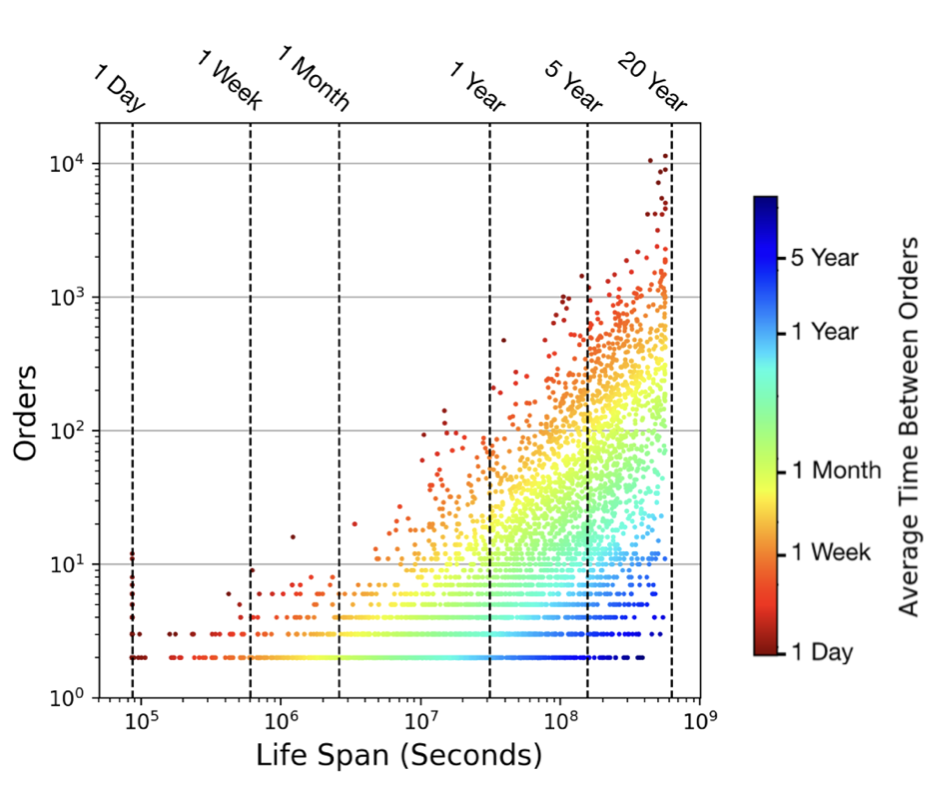}
\caption{\label{figs1}Customer orders and lifespan for the manufacturing company. Scatter plot of total number of orders, customer lifespan, and average time between orders (color bar, right) for customers for the period 11/1999 to 9/2017. Each dot represents one customer. Lifespan is the time between the first and last recorded order for each customer. Dashed lines (labels, top) delineate selected time periods. Axes and color bar are log scale.}
\end{figure}

\begin{figure*}[ht!]
\centering
\includegraphics[width=0.9\textwidth]{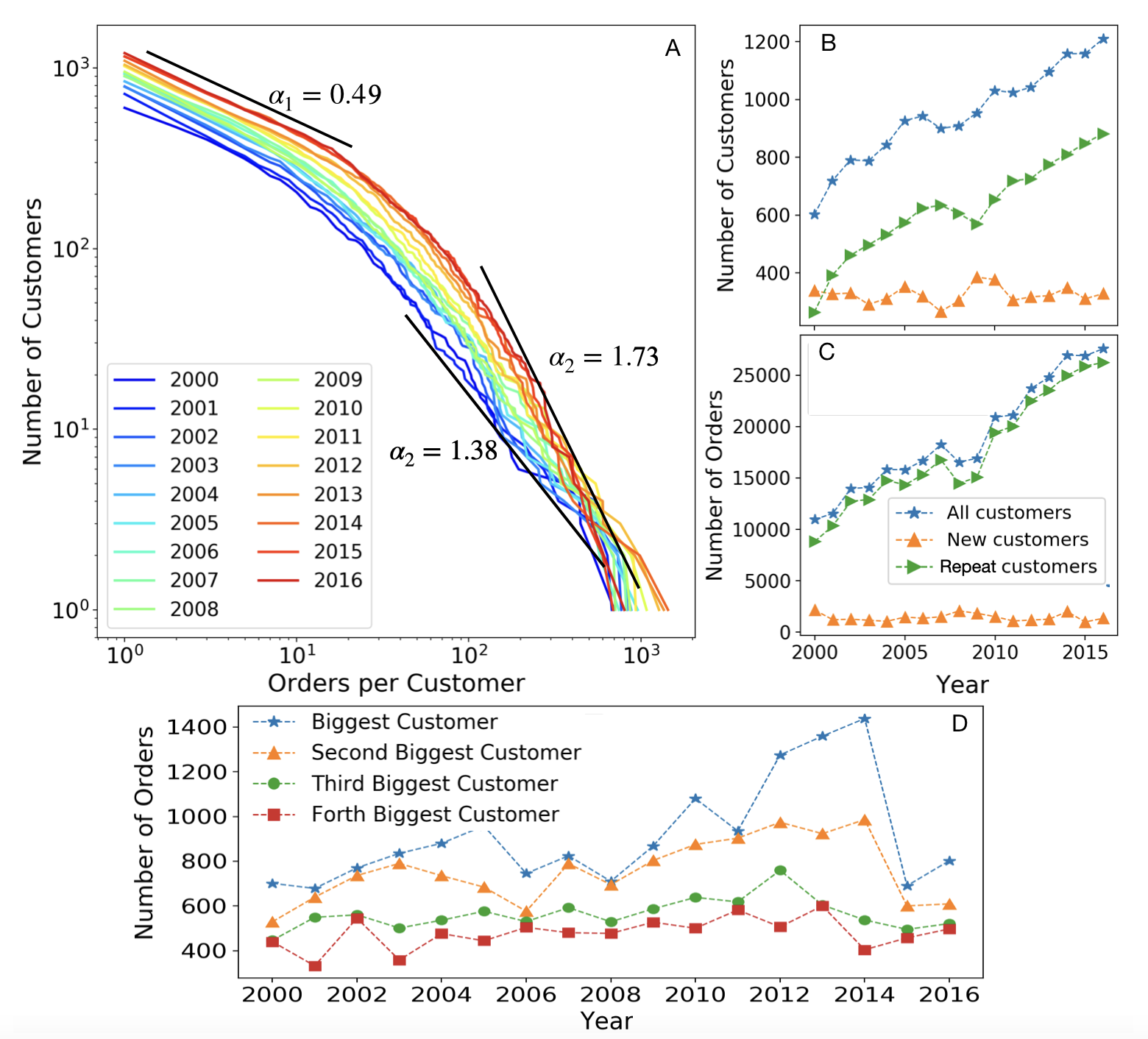}
\caption{\label{figs2}
Yearly power-law behavior of customers for the manufacturing company. (A) Number of customers $vs.$ orders on log-log axes, with colors representing years. (B-C) Number of customers and orders per year, respectively. Symbols denote all, new, or repeat customers (key, inset). (D) Number of orders per year for the four largest customers, indicated by symbols (key, inset).}
\end{figure*}

The behavior of customers ordering products from a company is quite heterogeneous. Figure \ref{figs1} shows a scatter plot characterizing customers for a midsized manufacturing company by total number of orders, customer lifespan, and average time between orders, with each dot representing one customer. The average time between orders is depicted on a color scale, with red being the most frequent (days) and blue being the least frequent (years). Customer lifespan is the time between the first and last recorded order, ranging from one day to the entire 18-year period. The total number of orders per customer ranges from 1 to over 10,000. Customers with few total orders (bottom of the graph) are the most heterogeneous, having a wide range of lifespans and times between orders. Meanwhile, customers with the highest number of orders (upper right corner) all have long lifespans, suggesting that a certain number of years is required to become a very large customer. We examined the yearly distribution of customers for the manufacturing company, counting the number of customers N(O) that place more than O orders (Fig. \ref{figs2}A). The distribution follows a power-law $N(O)\sim O ^{-\alpha}$, which includes two regimes with different $\alpha$ values. In the first regime, $\alpha_1= 0.49 \pm 0.02$, which is consistent for all years. In the second regime, $\alpha_2$ equals $1.38\pm 0.08$ from 2000-2013 and $1.73\pm 0.02$ from 2014-2016. The yearly curves are similar in shape but move outward over time, showing that the number of customers and orders is almost continuously growing across all scales of orders (Fig. \ref{figs2}B-C). Changes in the tails of the curves for 2014-2016 are largely due to operational changes, which disproportionately affected the number of orders of large customers (Fig. \ref{figs2}D).

\section*{II. PROBABILITY OF CUSTOMER PARAMETER SPACE}
Up to this point, we have explored how the distribution of customer orders follows a power law, and we have fit the customer order data to a sigmoidal curve. Now, we utilize probability functions of the sigmoidal parameters to formulate an analytical model that can tell us the probability of having a customer with a specific set of parameters. We then validate our results by comparing the distribution from our analytical model with the distribution from the customer data.

\subsection*{A. Probability Function}
In plotting the parameter spaces, we used four parameters: start time, inflection time, slope and amplitude. These parameters are not independent of each other. In fact, inflection time, slope and amplitude are related to each other through the sigmoidal curve. So, there are only three parameters that are fully or partially independent from each other. As long as we are able to define the probability function using three parameters, we can use it to find the probability functions for any other combination of three parameters. As start time $t_0$ is an independent parameter, the probability function should be a function of start time and any two other parameters. Here, we choose ln(slope) $m'$ and ln(amplitude) $A'$ and try to find an equation that shows the probability of having a customer that starts at time $t_0$ and reaches ln(slope) $m'$ with total ln(orders) $A'$ at time t, $p(t_0,m',A')$. The dependence of the probability on time $t$ is hidden in the borders of the start time $t_0$ parameter, $t_{0a}<t_0<t_{0b}$. If the parameters were independent of each other, the form of the probability function would be a simple multiplication of the probabilities of the parameters,
\begin{equation}
    p(t_0,m^{'},A^{'}) =p(A^{'}) p(m^{'})p(t_0)
\end{equation}
However, the parameters are not independent, and we
therefore define the probability as,
\begin{equation}
    p(t_0,m',A') =p(A'|m',t_0) p(m'|t_0)p(t_0)
\end{equation}
where the probability of the start time is independent of the other parameters and can be considered as a separate term. The probability of having a specific slope depends on the customer start time of ordering (Fig. 2). Therefore its probability is conditional and denoted by $p(m'|t_0)$, the probability of $m'$ given that $t_0$ has occurred. The probability of having a specific amplitude is also conditional on the start time of a customer and the slope at that time, $p(A'|m',t_0)$. 

\begin{figure}[ht!]
\centering
\includegraphics[width=0.9\linewidth]{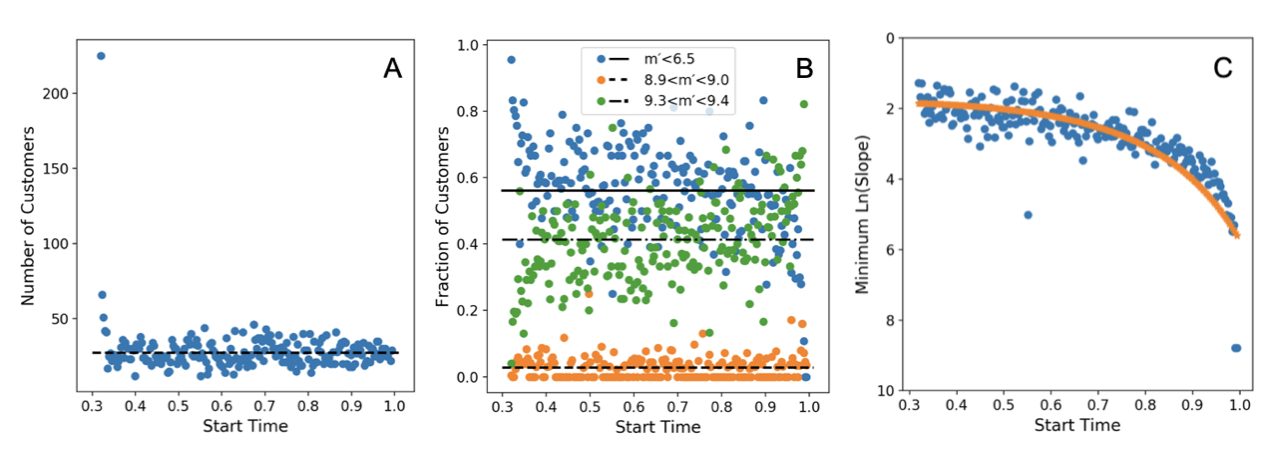} 
\caption{\label{figs3} Relationships among customer parameters. (A) The x-axis represents the first time of ordering and is rescaled from 0.32 to 1. (B) Fraction of customers of each type per start time. Green dots indicate customers with one or two orders with a ln(slope) of $9.3-9.4$. Orange dots are customers with less than 5 orders with a ln(slope) of $8.9-9.0$. Blue dots are all remaining customers with a ln(slope) less than $6.5$. (C) Minimum ln(slope) of all customers at each start time, with each dot representing one month. The fitted curve for all the points is shown in orange.}
\end{figure}

Probability of start at $t_0$: Start time represents the first time a customer shows up in the system and places an order. Fig. \ref{figs3}A shows a histogram of the number of customers starting in each month from $11/1999$ to $9/2017$. The x-axis represents start time, rescaled from $t_{0a}=0.32$ to $t_{0b}=1$. In this rescaled axis, the first month, 11/1999, is equal to 0.32, and the last month, 9/2017, is equal to 1. The nonzero value of $t_{0a}$ results from our introduction of 100 zero values at the beginning of the time series in order to ge a more reliable fit. The first month’s customer number is much larger than the other months’ because the company existed before 1999, while tracking of customer data did not begin until 11/1999, so the number of customers in the first month shows all the customers in the system up to that time. Apart from the first month, the histogram of customers represents an almost uniform frequency with an average value of 27 (red dashed line). This means that on average, 27 new customers start ordering each month. So we can say that customers are equally distributed over time based on their start time $t_0$ with a uniform probability,

\begin{eqnarray}
     p(t_0)= 
     \begin{cases} 
     \frac{1}{t_{0b}-t_{0a}} \quad & t_{0a}\leq t_0 \leq t_{0b} \\
     0 \quad  & \text{otherwise}
     \label{eqpt0}
     \end{cases}
\end{eqnarray}
where $t_{0a}=0.32$ and $t_{0b}=1$ represent the first and last months at $t=2017$. 

Probability of ln(slope) $m'$: The slope derived from the sigmoidal curve fit is sensitive to the lifespan of customers and the sequence of orders. As shown in Fig. 2 of the paper, many customers place only one or two orders and stay in the system less than one month. The ln(slope) for these customers is always around 9.3 -- 9.4. A few customers stay in the system less than 4 months and place up to 5 orders. The ln(slope) of these customers is around 8.9 -- 9.0. The rest of the customers, who have a ln(slope) of less than 8.9, have stayed with the company from 6 months to 18 years. Since there are only a few customers with ln(slope) values in the range $8.9>m'>6.5$ each month, we excluded them from our probability calculations for simplicity and considered the range $m'<6.5$ for repeat customers. Fig. \ref{figs3}B shows the fraction customers in each range of slopes versus the start time from 1999 - 2017. According to this figure, the fraction of customers of each type (slope range) is roughly constant as a function of start time $t_{0}$ and equal to $0.57$ for $m'_a <m'<6.5$, $0.03$ for $8.9 < m' < 9.1$, and $0.40$ for $9.3 < m' < 9.4$. We write the probability of having a customer with slope $m'$ and start time $t_{0}$ at time t as

\begin{eqnarray}
     p(m'|t_0)= 
     \begin{cases}
     \frac{0.57}{6.5-m'_a} \quad m'_a<m'<6.5 \\
   \frac{0.03}{9.1-8.9} \quad 8.9<m'<9.1 \\
   \frac{0.40}{9.4-9.3} \quad 9.3<m'<9.4
   \end{cases}
   \label{eqpm't0}
\end{eqnarray}

The sum of all these regimes is equal to one. The lower bound for the slope in the first regime of this equation depends on the start time of a customer. The slope of customers that start earlier can be chosen from a wider uniform distribution. The lower value is $m'_a = n_0 +e^{n_1t_0^{4}}$, where $n_0=0.5$ and $n_1=1.5$ for t = 2017. Fig. \ref{figs3}C shows the minimum slope of customers per start time, and the orange line shows the fitted line using the equation for $m'_a$. In the early years with less data, the values of $n_0$ and $n_1$ are relatively high, reaching approximately 2 and 10 around $t=2003$. However, in the subsequent years, these values stabilize around $0.5$ and $1.5$. Given the probability of the customer start time $p(t_0)$ in Equation \ref{eqpt0} and the conditional probability of the slope $p(m'|t_0)$ in Equation \ref{eqpm't0}, we are able to find the joint probability function of slope and start time of ordering,

\begin{equation}
    p(m^{'},t_{0})=p(m^{'}|t_{0})p(t_{0})
\end{equation}

This function represents the probability of having a customer that starts at time $t_0$ and reaches slope $m'$ at time $t$. 

\begin{figure}[ht!]
\centering
\includegraphics[width=0.7\linewidth]{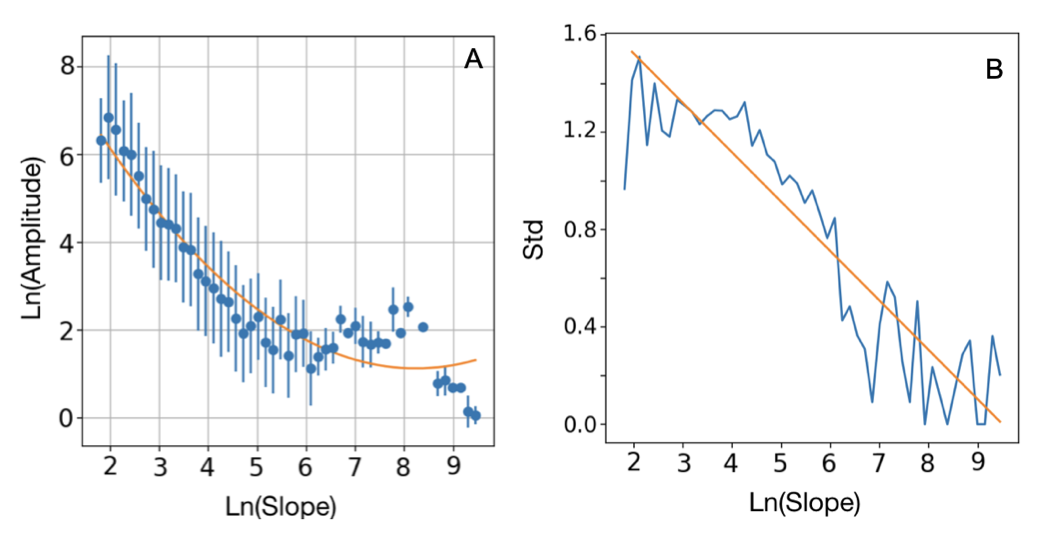}
\caption{\label{figs4} Relationship between amplitude and slope of customers. Panel (A) shows the mean amplitude (blue dots) as a function of the slope, with error bars representing the standard deviation of the amplitude. The orange curve is the second-order polynomial fit to the points. Panel (B) shows the standard deviation of amplitude versus slope, with the blue line representing the standard deviation and the orange line the linear fit.}
\end{figure}

\begin{figure}[ht!]
\centering
\includegraphics[width=0.9\linewidth]{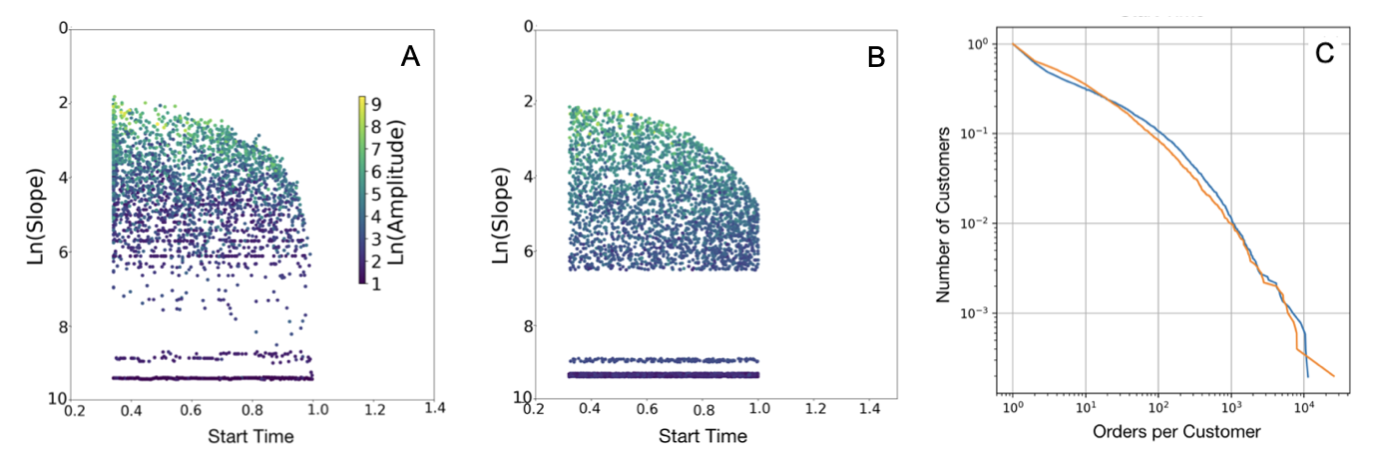}
\caption{\label{figs5} Panels (A) and (B) are parameter spaces for the customer data and the simulated data, respectively, with slope versus start time on the axes and amplitude in colors (key on upper right). Panel (C) depicts the distribution of number of customers versus orders per customer, with the customer data shown in orange and the simulated data in blue. The two distributions are in close agreement, indicating that the analytical model has generated a realistic distribution.}
\end{figure}

Probability of Amplitude $A'$: The amplitude a customer can reach has an inverse relation with the slope (Fig. \ref{figs4}A). We fitted a normal distribution to the relationship between the amplitude and the slope:

\begin{eqnarray}
     p(A^{'}|m^{'},t_{0})= 
     \begin{cases}
     \frac{1}{\sqrt{2\pi}\sigma_{1}} e^{-\frac{(A^{'}-\mu_{1})^{2}}{2\sigma^{2}_{1}}} \quad m^{'}_{a}<m^{'}\leq 6.5 \\
   \ \frac{1}{\sqrt{2\pi}\sigma_{2}} e^{-\frac{(A^{'}-\mu_{2})^{2}}{2\sigma^{2}_{2}}} \quad 6.5<m^{'}\leq 9.4
   \end{cases}
\end{eqnarray}
where the parameters for the first regime are given by $\mu_{1}=a_{0}m^{'2}+a_{1}m^{'}+a_2$ and $\sigma_1=b_{0}m^{'}+b_{1}$ (Fig. \ref{figs4}B), with $a_{0}=0.13$ , $a_{1}=-2.12$ , $a_{2}=9.86$, $b_{0}=-0.20$ , and $b_{1}=1.93$, and the parameters for the second regime are $\mu_{1}=1$ and $\sigma_{1}=ln(1.5)$. The effectiveness of the fitting using this parameterization is shown in Fig.  \ref{figs5}B. Large customers (ln(slope) $<$ 6.5) have a mean amplitude that is a good fit to the polynomial function, as shown in the figure. For large customers, the standard deviation of the amplitude decreases roughly linearly as the slope increases. Meanwhile, small customers (ln(slope) $>$ 6.5) have a mean amplitude that does not fit the polynomial function and instead fluctuates around 1.0. The error bar for small customers also does not follow the fitted line for large customers and instead fluctuates around $ln(1.5)$.

\subsection*{B. Validation of the Result}

We validated our analytical model by generating a sample population using the probability functions from the previous section. We use the equations for the probability of a customer with a given start time (Eq. 3), slope (Eq. 4), and amplitude (Eq. 6) to generate the values for Equation 2. The resulting parameter spaces for the customer data and the simulated data are shown in Fig. \ref{figs5}A and B, respectively. Next, we generated distributions of the number of customers versus orders for both data sets. There is close agreement between the customer (orange) and the simulated (blue) distributions (Fig. \ref{figs5}C). Overall, the tests indicate that our analytical model performs well and can be used to generate predictions of customer behavior.

\section*{IV.RESULTS FOR THE LEGISLATIVE DATA SET}

Legislative data refers to information related to the legislative process, including the activities and decisions of legislative bodies such as parliaments, congresses, or other similar institutions. We employed the lifepaths methodology to assess and forecast future legislative developments related to Per- and Polyfluorinated Substances (PFAS). This data encompasses a wide range of information about proposed, enacted, and rejected legislation, as well as the people and organizations involved in the legislative process. The LegiScan platform primarily displays the number of bills for the US and individual states. However, a more nuanced understanding of the driving forces behind legislative actions can be gleaned through visualizing essential keywords and named entities present in bills.

Legislation seldom materializes in isolation. Frequently, the introduction of bills is catalyzed by antecedent acts referenced within the text of the proposed bills. In the context of PFAS-related bills, legislators commonly invoke acts such as the Clean Air Act (CAA) and the Safe Drinking Water Act (SDWA). Both are pivotal federal laws in the US regulating air and water quality standards, granting authority to the Environmental Protection Agency (EPA) to establish standards safeguarding public health and welfare, as well as to regulate hazardous pollutant emissions. Consequently, acts, environmental agencies, associations, agencies, boards, various codes delineating minimum requirements (e.g., building codes or energy conservation codes), and numerous other named entities often serve as catalysts in the legislative process. Vigilance towards the dynamics of these entities may offer valuable insights into the potential introduction of future bills concerning PFAS.

The challenge of characterizing behaviors of terms (terms in this context means named entities) in documents arises from a fundamental property: complexity. Usage of terms in legal bills is a complex behavior, as authors who introduce bills attempt to maximize their influence independently. Time and influence limitations can impose constraints on usage of certain terminology, bound their lifetime, and make terms adopt nonlinear behaviors. Therefore, terms continuously enter and leave semantic spaces of legal bills over time. This complexity of terms at the individual level increases the complexity of the system in aggregate. 

\begin{figure}[ht!]
\centering
\includegraphics[width=0.5\linewidth]{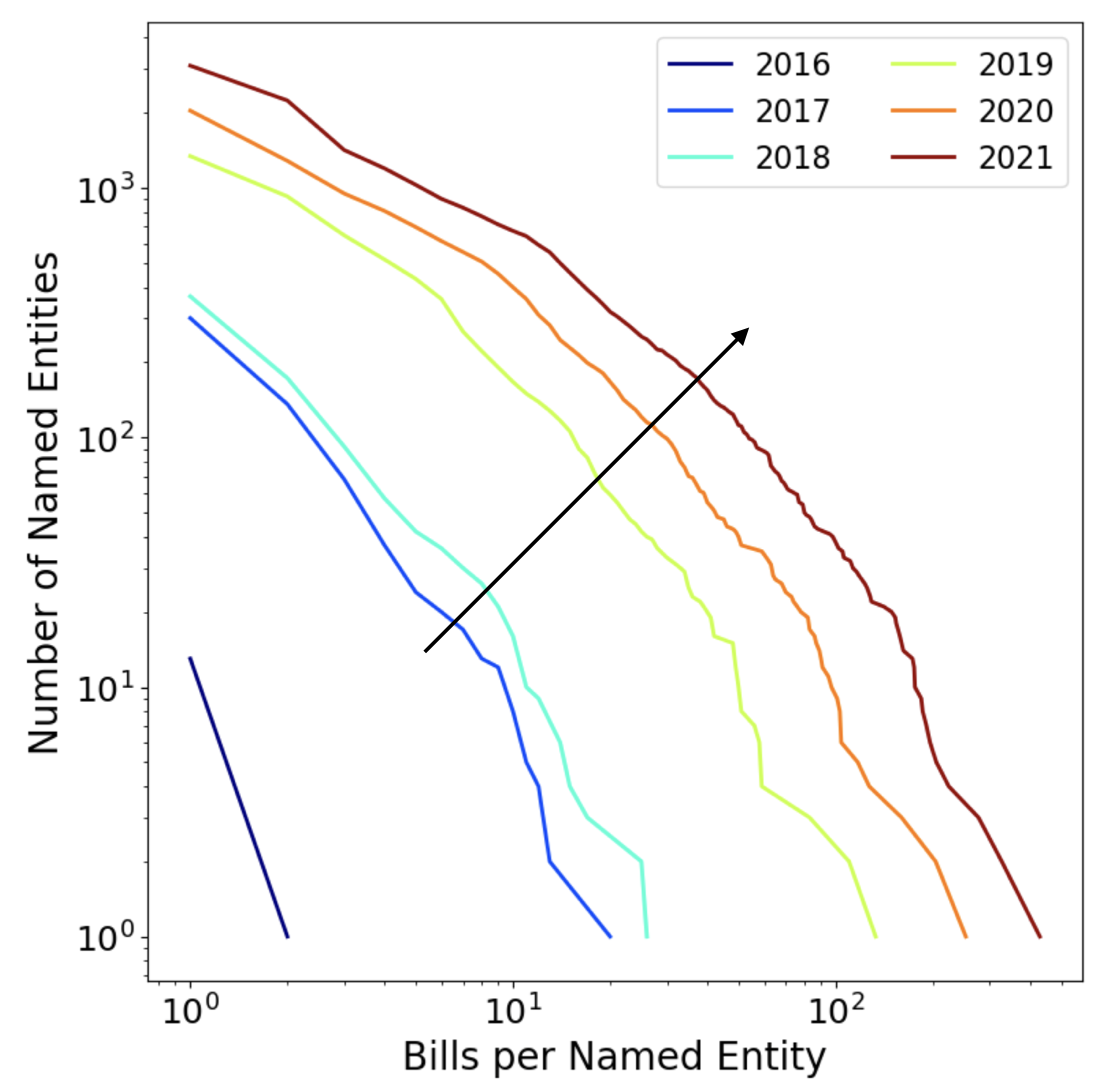}
\caption{\label{figs10}  Scaling behavior of named entities in relation to the number of bills concerning PFAS from 2016 to 2021. }
\end{figure}

Figure \ref{figs10} illustrates the scaling behavior of the cumulative number of named entities versus the number of bills from 2016 to 2021. We collected 3,100 named entities (NEs) from over 430 introduced bills. The data shows that, over the years, every NE appeared in at least one bill, while one NE appeared in all bills. The logarithmic representation of the axes indicates a power law behavior, with a noticeable change in the power at a specific number of bills. This behavioral shift represents the time limitation for introducing new bills for NEs over time. 

\begin{figure}[ht!]
\centering
\includegraphics[width=0.6\linewidth]{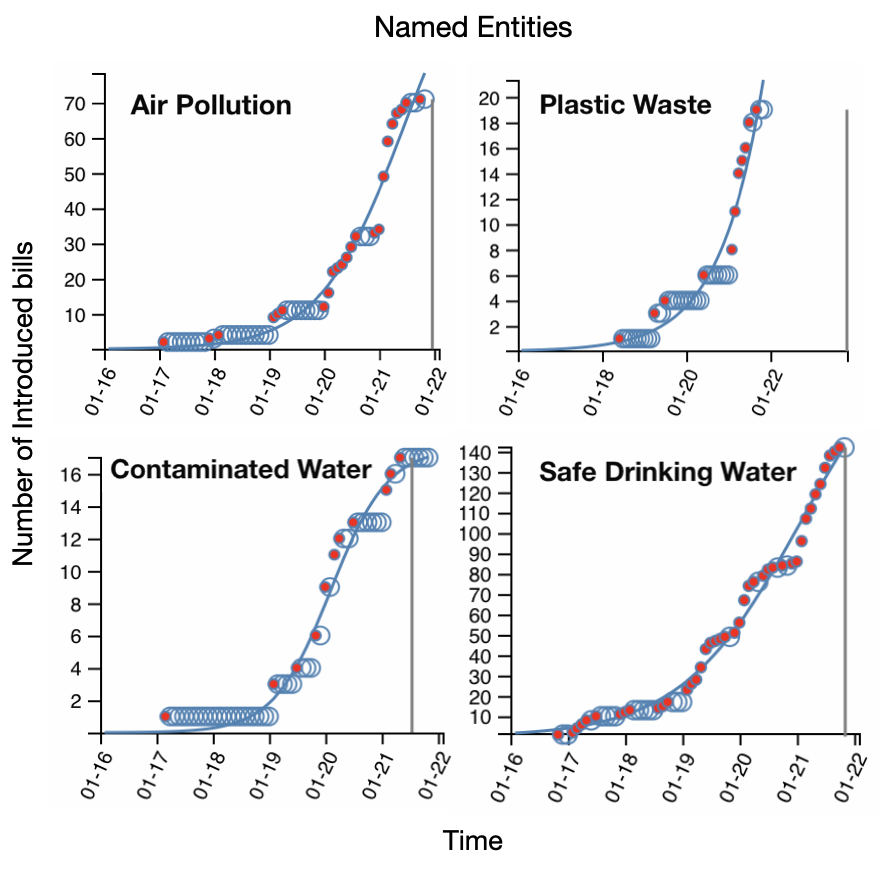}
\caption{\label{figs11} Sigmoid fitting pattern for four selected NEs. The red circles indicate the number of bills that include each NE across time steps. The blue circles replicate the number of bills in consecutive time steps to clarify the fitting process. The blue line represents the sigmoid fitted curve, while the black vertical line marks when the inflection point happens. }
\end{figure}

\begin{figure}[ht!]
\centering
\includegraphics[width=0.6\linewidth]{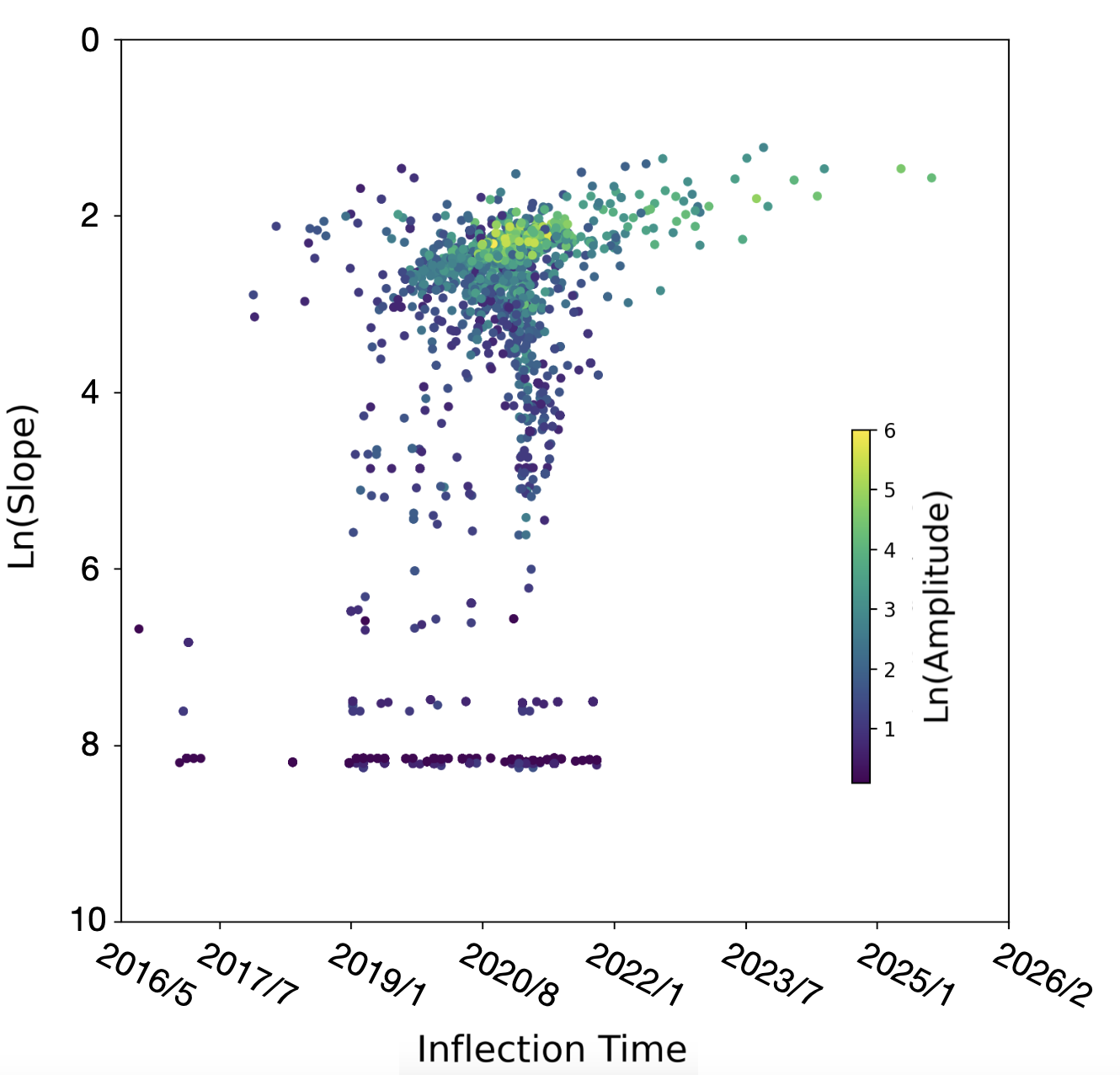}
\caption{\label{figs12} Sigmoid fitting parameter space analysis, inflection time, slope, and amplitude of named entities in the year 2020. }
\end{figure}

For named entities that gradually enter or exit the semantic spaces of legal bills related to PFAS, the term dynamics can be accurately modeled using a sigmoid function. Figure \ref{figs11} represents the fitting process to four examples of NEs including: Air Pollution, Plastic Waste, Contaminated Water, and Safe Drinking Water. When the fitting parameters are plotted collectively, they reveal patterns of behavior that are not discernible at the individual level. Figure \ref{figs12} presents the parameter space of NEs, including inflection time, slope, and amplitude, for the year 2020. NEs with low frequency and few associated bills are located at the bottom of the figure, characterized by a high slope value. Conversely, frequently occurring named entities that appear in numerous bills are positioned higher in the parameter space, exhibiting lower slope values. Emerging named entities are found in the top right of the figure, indicating an inflection time in the future. 

\begin{figure}[ht!]
\centering
\includegraphics[width=0.5\linewidth]{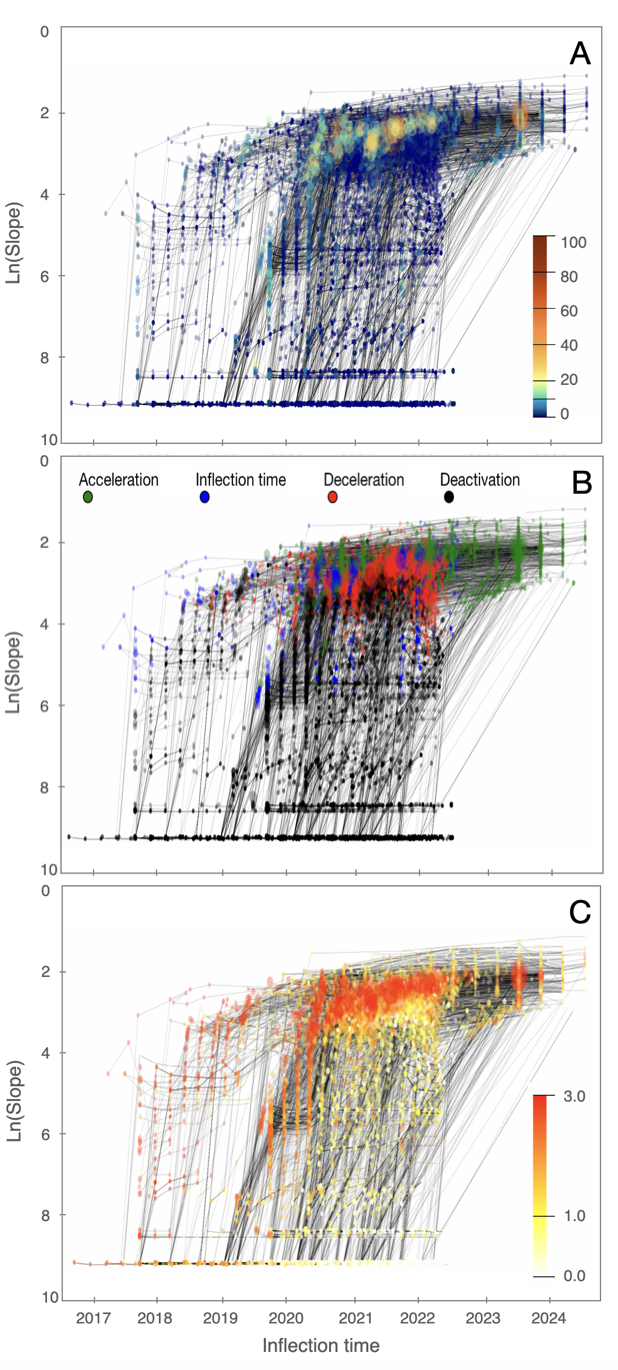}
\caption{\label{figs13}  Lifepath representations to assess and forecast future legislative developments related to PFAS. A) Represents numbers of bills introduced at each step. B) Represents lifetime statuses, Acceleration/Deceleration/Deactivation/Inflection Time, and C) shows which lifepaths are local and which ones are global (i.e., appearing in many states) using an entropic measure.}
\end{figure}

Figure \ref{figs13} represents the lifepaths of NEs over the years 2016 to 2022. Lifepaths merge graphs of parameter spaces from consecutive years to present the entire sigmoid analysis in a single frame. Lifepaths have three representations: 1) representation with numbers of bills introduced at each step, Fig. \ref{figs13}A; 2) representation with lifetime statuses: Acceleration/Deceleration/Deactivation/Inflection Time, Fig. \ref{figs13}B; and 3) representation showing which lifepaths are local and which ones are global (i.e., appearing in many states), Fig. \ref{figs13}C. At the beginning, the lifepaths have large logs of slopes; they are located at the bottom of the plot. We can classify the sigmoid curves into four phases: acceleration, at inflection point, deceleration, and inactivation or saturation phase, see Fig. \ref{figs12}. In the acceleration phase the time series is lagging behind the inflection point at the time of the analysis ( time of analysis $<$ predicted inflection time). We colored the accelerating portions of lifepaths green. In this phase, the number of bills is growing fast. When the predicted inflection time from the sigmoid fitting is equal or close to the time of the analysis, then the lifepath is at the inflection point and is marked blue. During this phase, the lifepaths transition from acceleration to deceleration. When a time series passes its inflection point (time of analysis $<$ predicted inflection time), it migrates to the deceleration phase (shown black in the plot). In this phase, the numbers of bills in lifepaths are growing slowly. When the time series reach their saturation points, bills stop being introduced, and the time series growths are deactivated. 

 Each representation of all lifepaths together shows in which part of the plot similar behaviors are accumulated. For example, examining statuses, users can see that accelerating behaviors happen when the slope values are between 2 and 4, and they extend in time far ahead of all other statuses. Deactivated statuses of named entities are shown in black; they form a sediment layer. The sediment layer occupies a large part of the plot. New lifepaths start growing from the sediment layer. Some named entities manage to reach Acceleration, Inflection Time and Deceleration stages, others remain in the sediment (aka deactivated) all the time. Representations 1 and 3 (Fig. \ref{figs13}A and C) partially explain why this is happening. First, we can see from plot A that in the top part of the plot there are many NEs with a large number of bills at each step. Second, many NEs are used in legislation of multiple US states, as evaluated using Shannon entropy measure, when they start accelerating or decelerating. 
 \begin{equation}
     H(X) = - \Sigma_{i=1}^{n} P(x_i) \ln P(x_i)
 \end{equation}
Here, $H(X)$ represents the entropy of term $X$, where $P(x_i) = y_i/\Sigma_i(y_i)$ is the probability of term X occurring in State $i$. The term $y_i$ accounts for total occurrences of the term across all states. Entropy measures how influential terms are across bills in different states.

In conclusion, these collective representations of PFAS-related entities in the legislative system illustrate how various interested parties, boards, organizations, and associations continually introduce bills to ban PFAS products. Tracing these entities over time allows analysts to observe public activity regarding product safety regulation. In some states, bans are initiated by a few entities, while in others, there are many active participants, with varying levels of activity. Some entities appear sporadically in the legislative system and do not reach the upper part of the parameter space (Fig. \ref{figs12}). Entities in the acceleration phase each year are the primary drivers in the legislative system across all states in the US and are likely to remain active in the coming years. Entities that have reached their inflection point are those that have driven new bills until recently but are now slowing down. Finally, entities in the deceleration phase have garnered attention in past years, with most of the necessary bills already introduced.